# Effective interaction model for coupled magnetism and phase stability in bcc Fe-Co systems

**Van-Truong Tran[*], Chu-Chun Fu[‡], and Anton Schneider**

DEN-Service de Recherches de Métallurgie Physique, CEA, Université Paris-Saclay, F-91191 Gif-sur-Yvette, France

Emails: [*]vantruongtran.nanophys@gmail.com and [‡]chuchun.fu@cea.fr

## Abstract

We present an *ab-initio*-based effective interaction model (EIM) for the study of magnetism, thermodynamics, and their interplay in body-centered cubic Fe-Co alloys, with Co content from 0 to 70%. The model includes explicitly both spin and chemical variables. For the former, a Heisenberg formalism is adopted. But, the spin magnitude of each Fe atom varies according to its local chemical environment, following a simple rule determined by density functional theory (DFT) calculations. The proposed model is able to describe precisely the ground-state magneto-energetic landscape of both chemically ordered and disordered Fe-Co systems, as given by DFT and experiments. In combination with on-lattice Monte Carlo simulations, it enables an accurate prediction at finite temperatures. In particular, the Curie point and the chemical order-disorder (B2-A2) transition temperature are accurately predicted, for all the concentrations considered. A strong dependency of the chemical transition temperature on the magnetic configuration is evidenced and analyzed. We also suggest a more important effect of magnetic rather than vibrational entropy on the chemical transition. However, this transition is not affected by a commonly accessible external magnetic field.

## I. Introduction

Body-centered cubic (bcc) Fe-Co alloy in the ferromagnetic (FM) state has been found as the highest magnetic system among all known alloys [1]. Additionally, it is demonstrated experimentally that the Fe-Co based systems exhibit high Curie temperatures, up to about 1250 K [2,3]. All these outstanding features make this alloy extremely attractive for various magnetic applications, particularly for the construction of strong magnets, to be used for several (from commercial to military) applications [1,4].

Due to the appealing magnetic properties, Fe-Co systems have been studied extensively in order to better understand and to modulate such properties. Theoretically, density functional theory (DFT) calculations are widely used for the study of magnetic and energetic properties [5,6,15–19,7–14]. This method has shown to successfully predict properties in good agreement with low-temperature experimental data [15–19]. However, its application is limited to rather small simulation cells, and more importantly, it does not provide a direct prediction of magneto-thermodynamic properties as functions of temperature.

Considering semi-empirical simulations, magnetic-interaction models within the Ising or the Heisenberg formalism are broadly employed in combination with the mean-field theory or Monte Carlo (MC) simulations. Such approaches allow us to explore efficiently the temperature-dependence of magnetic properties of large systems composed of atomic spins. Indeed, the Ising or Heisenberg models were developed in previous works for Fe-Ni [20], Fe-Cr [21], Fe-Mn [22,23], Fe4N [24], Fe-Al [25,26] and Fe-Co systems [27,28].

To go further, and in order to account for both magnetic and chemical degrees of freedom in a magnetic alloy, other models including explicitly these two variables have been proposed, for example, by early studies of Tahir-Kheli *et at.* [29] and Sanchez *et at.* [30]. The magnetic-interaction part commonly adopted the Ising or the Heisenberg form. Such models were parameterized for various alloys such as Fe-Ni [20,31,32] and Cu-Al-Mn [33]. Recently, more sophisticated models called magnetic cluster expansion (MCE) models have been developed by Lavrentiev *et. al.* for Fe-Cr, Fe-Ni, Fe-Cr-Ni systems [34–40], where the magnetic part has a Landau-Heisenberg form, in order to include both longitudinal and transversal excitations of spins. In addition, these MCE models generally go beyond the simple pair interactions. Thus, coupled spin-atom Monte Carlo simulations using such models can be highly computationally demanding. So far, they have been mostly applied for the study of magnetic properties with a fixed chemical configuration [34–38], while much less for thermodynamic properties [39,40].

Regarding the case of bcc Fe-Co alloys, various interaction models containing both spin and chemical variables have been proposed so far, by Pierron-Bohnes *et. al.* [41], and more recently by Prasad *et. al.* [42] to investigate thermodynamic properties considering the magnetism. These two interaction models adopted the Ising-like magnetic interactions, and were treated within the mean-field approach. The latter approach provided a description in good agreement with experimental transition temperatures, but it presented magnetic exchange-coupling parameters dependent on the alloy concentration. However, please note that, as revealed by experiments [41] and in previous DFT studies [43], the magnitude of Fe magnetic moment depends sensitively on the chemical long and short range orders. Therefore, magnetic interactions in Fe-Co cannot be correctly captured by the conventional Ising and Heisenberg models.

In this work, we aim at proposing an effective interaction model (EIM) for the study of magnetism, chemical-phase stability and their coupling in bcc Fe-Co structures. This EIM contains two parts: (i) a non-magnetic part describes the chemical interaction between atoms; and (ii) a magnetic-interaction part with a Heisenberg formalism. But, the spin magnitude of each Fe atom varies according to its local chemical environment, following a simple rule, determined by our previous DFT calculations [43]. In order to keep the model as simple as possible, and efficient for Monte Carlo simulations, we only include pair-interaction terms, at variance with the MCEs. Also, unlike the previous approach in Ref. [44], a unique set of interaction parameters is determined for the whole range of alloy concentration considered (from 0 to 70% Co). The EIM is therefore highly transferable. In addition, we also elucidate correlations between the magnetic state and the relative stability of chemically ordered and disordered phases for various temperatures and Co concentrations.

The paper is organized as follows: the proposed effective interaction model for bcc Fe-Co systems is described in Sec. II, where we first present the Hamiltonian (Sec. II 1) and then give the details about the parameterization (Sec. II 2). The used DFT and Monte Carlo simulation methods are given in Sec. III, together with the expression of various considered order parameters. Sec. IV is devoted to some results from the EIM and discussions. First, validation of the model is shown in Sec. IV 1. Then, finite-temperature magnetic and thermodynamic properties are predicted in Sec. IV 2, and, effects of an external magnetic field are discussed in Sec. IV 3. Finally, conclusions are given in Sec. V.

## II. The effective interaction model (EIM)

## 1. The model

The effective interaction model is written as

$$H = \frac{1}{2}\sum_i\sum_j V_{ij}\sigma_i\sigma_j + \frac{1}{2}\sum_i\sum_j J_{ij}\overrightarrow{M_i}\overrightarrow{M_j}, \qquad (1)$$

here the first part accounting for non-magnetic contributions, describes chemical interactions between atoms. $V_{ij}$ are the chemical pair interaction parameters between *i*-th and *j*-th atoms and $\sigma_{i(j)}=1$ if the lattice site is occupied by an atom. The second part of Eq. (1) corresponds to the Heisenberg model with $J_{ij}$ being the exchange-coupling parameter for the magnetic moments of *i*-th and *j*-th atoms. However, unlike the traditional Heisenberg model in which the amplitude of an atomic moment is fixed, the magnetic part of the model (1) takes into account the longitudinal spin fluctuation due to the effect of the local environment as observed by the DFT calculations [43], i.e., for a Fe atom, the magnetic moment can be described by a vector in the spherical coordinate system $\overrightarrow{M_{Fe}} = (M_{Fe}, \varphi, \theta)$ with $\varphi$ and $\theta$ being polar and azimuthal angles, respectively. The amplitude $M_{Fe}$ is defined as

$$\begin{aligned}M_{Fe} = P_0 &+ A_1 \times N_1 + A_2 \times N_1^2 + A_3 \times N_1^3 + B_1 \times N_2 + B_2 \times N_2^2 + B_3 \times N_2^3 \\ &+ C \times N_1 N_2 + D_1 \times N_1^2 N_2 + D_2 \times N_1 N_2^2\end{aligned} \qquad (2)$$

This equation is obtained from the analyses of the effect of the local environment on magnetic moments in bcc Fe-Co systems that are presented in detail in Ref. [43]. In Eq. (2), $N_1$ and $N_2$ are the numbers of Co atoms in the first and second neighboring shells of the central Fe atom. The fitting parameters $P_0$, $A_1$, $A_2$, $A_3$, $B_1$, $B_2$, $B_3$, $C$, $D_1$ and $D_2$ were taken from Ref. [43].

For Co atoms, it is observed from the DFT calculations [43] and experiments [44] that the average moment of Co atoms stays almost constant at the value of 1.76 $\mu_B$. Hence, the magnetic moment vector of a Co atom writes

$$\overrightarrow{M_{Co}} = (M_{Co} = 1.76, \varphi, \theta) \qquad (3)$$

Thus, the model allows the magnetic moment vector of atoms to vary in 3D space, making spin polarization fully compatible with the non-collinear treatment.

It is worth noting that we assume the amplitude of a moment being insensitive to temperature. Although there is no clear experimental evidence to support this assumption, thermal longitudinal fluctuations of spins in these systems were suggested to be negligible in previous theoretical studies [45,46].

## 2. Parameterization

The model (1) was first parameterized for $J_{ij}$ of the magnetic part by using a set of energy differences between two configurations of distinct magnetic states but with the same atomic configuration, i.e. $\Delta E$ which contains only the magnetic contribution.

$$\Delta E = \frac{1}{2} \sum_i \sum_j J_{ij} . \Delta\left(\vec{M_i}\vec{M_j}\right) \tag{4}$$

By performing DFT calculations for different bcc Fe, Co, and Fe-Co structures, a set of Eq. (4) was constructed, and a least squares fitting was performed to obtain the $J_{ij}$.

After obtaining the exchange-coupling parameters $J_{ij}$, the chemical interaction parameters $V_{ij}$ were deduced by considering the mixing enthalpies of special quasi-random Fe-Co structures. These mixing enthalpies are defined as [47]

$$E_{\text{mix}} = \varepsilon_{FeCo} - X_{Co}.\varepsilon_{Co} - X_{Fe}.\varepsilon_{Fe}, \tag{5}$$

in which $\varepsilon_{FeCo}$ is the energy per atom of the Fe-Co structure and $\varepsilon_{Co}, \varepsilon_{Fe}$ are the energies per atom of the pure bcc Co, Fe systems, respectively. $X_{Co}$ and $X_{Fe}$ are the atomic fractions of Co and Fe in the Fe-Co structure.

In the EIM Hamiltonian (1), the magnetic interactions (via $J_{ij}$) are considered up to the 5-th neighboring shell of each atom, which is demonstrated to be sufficient for the precise description of the magnetic properties [36]. Thus, it requires to know of 15 $J_{ij}$ parameters corresponding to Fe-Fe, Co-Co and Fe-Co pair interactions. For the chemical part, we found that a cut-off of $V_{ij}$ at the second shell is enough to capture important chemical interactions between atoms. And thus, there are in total of 6 $V_{ij}$ parameters.

In summary, all $V_{ij}$ and $J_{ij}$ parameters of the Hamiltonian (Eq. 1) were defined by fitting to the DFT data of a variety of bcc Fe, Co, and Fe-Co structures and in different magnetic states to ensure that the model has the capacity to describe physics at both 0 K and high temperatures. The set of parameters is given in Table I.

## III. First-principles calculations and Monte Carlo simulations

### 1. Density functional theory methodology

The DFT calculations were carried out using plane-wave basis sets and projector augmented wave (PAW) potentials, as implemented in the Vienna Ab-initio Simulation Package (VASP). [48–51]

The GGA-PBE functional was utilized [52]. The calculations are spin polarized within the collinear approximation. The criterion of $10^{-6}$ eV was employed in electronic self-consistent loops. The studied structures were fully relaxed until the forces on atoms being less than 0.01 eV/Å and stresses on the cells being smaller than 5 kbar. The plane-wave energy cut-off of 400 eV was chosen to have a good convergence of energetic, magnetic and structural properties. The Methfessel-Paxton smearing function with a width of 0.1 eV was used. The Brillouin zones were sampled by the Monkhorst-Pack method [53] with a k-mesh of 16×16×16 for the simple cubic unit cell and an equivalent grid for a supercell: for example, a k-mesh grid of 4×4×4 was used for a supercell of 4×4×4 containing 128 atoms.

In this study, special quasi-random structures (SQS) were employed in DFT calculations to mimic disordered systems. All the SQSs and the partially ordered structures of bcc Fe-Co are represented by 128-atom supercells, while the cells used for the ordered phases of Fe-Co and pure bcc Co, Fe are case dependent.

### 2. Monte Carlo simulation setup

In this study, MC simulations were performed using the EIM (Eq. 1) for studies at finite temperatures. Two types of equilibrations were considered with MC simulations: (i) Spin equilibration at a given atomic configuration, which is referred as spin MC simulations; (ii)

Coupled magnetic and chemical equilibrations (barely spin-atom MC) which allow both spin variations and the exchange of position between pairs of atoms.

Both spin variations and atom exchanges are controlled by the Metropolis algorithm assuming the Boltzmann statistics.

In MC simulations, we used bcc systems of 16000 atoms (20×20×20 supercells) which were checked to be sufficiently large to ensure the convergence of energy and magnetization. Such a size of bcc systems for MC simulations was also used in previous studies [36,37,54].

For the spin equilibration, about $2.0\times10^8$ MC steps (number of spin variation attempts) were used to get convergence at each temperature.

Regarding the spin-atom equilibrations, two-stage simulations were carried out: (i) a spin equilibration was performed for the initial atomic configuration, (ii) a spin-atom equilibration runs with about $1.5\times10^6$ to $3.0\times10^6$ MC atom exchange steps. After each exchange attempt of a pair of atoms, about $10^3$ spin MC steps were performed some globally in the supercell and others locally to ensure the equilibrium state of the spin configuration corresponding to the new atomic configuration.

It is worth mentioning that MC simulations were performed at a given temperature to find the corresponding equilibrium state, and then the simulation at the next temperature starts from the converged spin and atomic configurations of the previous temperature. This procedure helps to speed up the reach-to-equilibrium of the system at successive temperatures.

## 3. Order parameters

In this study, several concepts are used to quantify the magnetic and chemical order of the studied systems. Chemical short-range order (chem-SRO) is widely employed and particularly useful to characterize solid solutions where chemical long-range ordering (chem-LRO) is missing. However, the temperature evolution of chem-LRO is used to monitor the chemical phase transition from an ordered to a disordered state or vice versa. Similarly, magnetic short-range order (mag-SRO) and long-range order can also be defined to characterize magnetic structures. The latter is actually the magnetization for ferromagnetic systems.

*Chemical short-range order:*

To determine the chem-SRO parameters of an alloy composed of two species A and B, first, the local arrangement of atoms is defined by the Warren-Cowley formula: [55,56]

$$\alpha_{B_i}^n = 1 - \frac{Z_{A-B_i}^n}{Z^n(1-X_B)}, \qquad (6)$$

here $\alpha_{B_i}^n$ is the parameter calculated for the n-*th* neighboring shell of a B atom, $Z_{A-B_i}^n$ is the number of A atoms in the n-*th* neighboring shell of the B atom. $Z^n$ is the total number of atoms in the n-*th* shell. In bcc structures, $Z^1 = 8$, $Z^2 = 6$ and so on. $X_B$ is the global concentration of the B atoms in the alloy. The chem-SRO parameter (on the B atoms) is defined as the average of the individual $\alpha_{B_i}^n$ over all the B atoms:

$$\langle \alpha_B^n \rangle = \frac{\sum_{i=1}^{N_B} \alpha_{B_i}^n}{N_B} \qquad (7)$$

*Chemical long-range order:*

Chem-LRO is simply defined as follows [42]

$$LRO_A = \left| c_1^A - c_2^A \right|, \qquad (8)$$

where $LRO_A$ is the chem-LRO of the species A, $c_1^A = N_1^A / N_1$ and $c_2^A = N_2^A / N_2$ are the concentrations of the A atoms in two inequivalent sub-lattices 1 and 2 of the bcc crystal structure, respectively. $N_1 = N_2 = N/2$ where $N$ is the total number of sub-lattice sites. It is also worth noting that $LRO_B = LRO_A$.

*Magnetic short-range order:*

We define the local magnetic ordering parameter of a Fe atom as follows

$$\begin{cases} \beta_{Fe_i-Fe}^{\ n} = \sum_{j=1}^{N_{Fe_i-Fe}^{\ n}} \frac{\left(\overrightarrow{M_{Fe_i}}\overrightarrow{M_{Fe_j}}\right)/\left(\left|\overrightarrow{M_{Fe_i}}\right|\cdot\left|\overrightarrow{M_{Fe_j}}\right|\right)}{N_{Fe_i-Fe}^{\ n}} \text{ if } N_{Fe_i-Fe}^{\ n} \neq 0 \\ \beta_{Fe_i-Fe}^{\ n} = 0 \text{ if } N_{Fe_i-Fe}^{\ n} = 0 \end{cases} \quad (9)$$

where $\beta_{Fe_i-Fe}^{\ n}$ is the magnetic ordering parameter of Fe$_i$. It represents the spin-spin correlation between the $i$-th Fe atom and its Fe neighboring atoms in the $n$-th shell. $N_{Fe_i-Fe}^{\ n}$ is the number of Fe neighboring atoms in the $n$-th shell of the atom Fe$_i$. Mag-SRO of Fe-Fe interactions is then defined as

$$\beta_{Fe-Fe}^{\ n} = \sum_{i=1}^{N_{Fe}} \frac{\beta_{Fe_i-Fe}^{\ n}}{N_{Fe}}, \quad (10)$$

with $N_{Fe}$ is the number of Fe atoms in the system.

A similar definition of the mag-SRO is applied for Fe-Co and Co-Co magnetic couplings.

For example, the ordered B2 structure has chem-SRO equal to -1 and 1 for the first and second shells, respectively, while its chem-LRO is equal to 1. The ideal SQSs have both chem-SRO and chem-LRO equal to zero. While the ferromagnetic systems have mag-SRO equal to 1 for every shell, and the perfectly paramagnetic (PM) ones have mag-SRO equal to 0.

## IV. EIM-based results and discussions

### 1. Validation of the effective interaction model

The quality of the model (Eq. 1) parameterization is firstly verified by checking its capacity to reproduce the energy difference ($\Delta E$) between any two distinct magnetic states of a same atomic configuration. A comparison between the $\Delta E$ obtained by the model and from DFT is displayed in Fig. 1. Most of the data points follow the diagonal line, showing an overall accuracy of the magnetic part. Only a couple of points deviate from the diagonal line, which indicates an over-estimation by the model of the energy of a few high-energy magnetic states such as the anti-ferromagnetic and certain disordered-magnetic systems.

Then, the model including both the magnetic and chemical parts is validated by comparing the mixing enthalpies of bcc Fe-Co from DFT and the model for 0 to 70% Co. To this end, we used the same DFT predicted atomic magnetic moments for the EIM. It is shown in Fig. 2 that the mixing enthalpies of the disordered structures (SQSs) by the model (green symbols) are in very good agreement with the DFT data (black symbols). Moreover, the results for the B2 structure are in excellent agreement between the two methods. This ensures the reliability of the model on thermodynamic properties of the Fe-Co alloys. As can be seen, both the DFT and the model exhibit negative mixing enthalpies of Fe-Co structures, which demonstrate a strong mixing tendency of Fe and Co atoms. Additionally, the B2 phase is more stable than the disordered one at 0 K.

Please note that in this paper, the atomic percentage is used to denote the composition of an alloy, and for the sake of simplicity, the prefix "at." is not explicitly written.

Concerning the magnetic properties, the model can be further validated by combining with spin MC simulations to obtain the magnetization of the Fe-Co structures at very low temperature (for example, at 1 K) and comparing them with the DFT results at 0 K. In Fig. 3, the average moment of each species Fe and Co as well as the total magnetization are shown for the disordered Fe-Co structures, which are often addressed as the A2 phase in the phase diagram. As can be seen, the results obtained by the model are in very good agreement with the DFT ones. As observed from Fig. 3, the average moment of Fe atoms increases remarkably from around 2.2 $\mu_B$ to 2.5 $\mu_B$ in the presence of even less than 20% Co, then it grows slowly up to a saturated value of 2.6 $\mu_B$ and remains almost constant from 50% Co. On the other hand, the magnetic moment of Co only varies slightly with the change of the alloy concentration and remains around 1.76 $\mu_B$. These results are fully consistent with the experimental observations. [44]

## 2. Prediction of finite-temperature properties

The use of the EIM in MC simulations allow us to predict a variety of properties at finite temperatures. In this section, several magnetic and thermodynamic properties of bcc Fe-Co systems are examined and discussed.

## 2.1 Magnetic and thermodynamic properties at a given atomic configuration

### 2.1.1 Magnetic transition and the Curie temperature

The magnetization of the disordered Fe-Co structures at high temperatures is predicted by the EIM and MC and shown in Fig. 4(a), while the average moments of Fe and Co species are represented in Fig. 4(b). As can be observed, the average moments of Fe and Co drop at high temperatures, leading to a lower magnetization. Such results could be understood as at high temperatures spins fluctuate strongly and the Fe-Co system transforms from the ferromagnetic state to the paramagnetic one.

By quantifying the magnetization as a function of temperature, the Curie temperature $T_c$ of a system can be determined when a sharp drop of magnetization is observed.

In Fig. 5, the magnetization curves are shown for bcc Fe, bcc Co, and bcc Fe-Co B2 and A2 at 50% Co. At first glance, within this model, the Curie points of the bcc Fe and Co systems are very close to each other and around 1100 K. While, the Fe-Co structures exhibit higher Curie temperatures, and the $T_c$ of the B2 is predicted to be larger than that of its A2 counterpart.

The determination of the Curie temperature may involve a significant uncertainty, because, due to the finite-size of the simulation box, the obtained magnetization does not drop sharply to zero at $T_c$. Several methods have been proposed, either using the inflection point of the magnetization-temperature curve [57,58] or the peak of the temperature evolution of the heat capacity [54,59]. However, these methods require a very dense sampling of the magnetization around the magnetic transition, therefore, computationally demanding. Alternatively, the fitting method using a power law has been widely used to determine $T_c$ [60–62]. All the data points of the magnetization before $T_c$ are less affected by the finite-size effect and can be used to fit the following expression:

$$m(T) = \frac{M(T)}{M(0)} = \left(1 - \frac{T}{T_c}\right)^{\beta}, \qquad (11)$$

where, $M(0)$ and $M(T)$ are the magnetizations at 0 K and the temperature $T$, respectively. A non-linear fitting can be used to define $\beta$ and $T_c$.

By using this power law (Eq. 11), $T_c$ = 1065 K is determined for pure bcc Fe in very good agreement with the experimental value of 1043 K [63]. Although the bcc phase of Co does not exist in nature, the model predicts that $T_c$ of this system is about 1077 K, which is lower than the $T_c$ (1400 K) of the fcc Co system [63].

For Fe-Co systems, $T_c$ are obtained to be respectively 1316 K and 1461 K for the 50% Co A2 and B2 structures. Similar calculations were also performed for other Co concentrations. The results allow us to obtain the magnetic phase diagram of bcc Fe-Co alloys (Fig. 6). The experimental data extracted from Ref. [64] containing data from Bozorth [63], Hansen [65] and Forrer [66] were also added for a comparison. However, please note that these experimental Curie temperatures may not come from direct measurements, but obtained by data extrapolations [66]. They roughly coincide with the bcc-fcc transition temperatures for a large range of Co concentrations. The Curie points obtained by our model via MC simulations are only slightly higher than these experimental data, and fully compatible with the equilibrium phase diagram, in which the bcc (B2 and A2) phase has a FM state while the fcc phase presents a PM state. Our predicted $T_c$ is found to increase significantly, from around 1060 K to above 1220 K, in the range from 0 to 20% Co, then it increases to the maximum value of about 1340 K at 40% Co before decreasing.

### 2.1.2 Mixing enthalpies and magnetic moments at finite temperatures

Mixing enthalpy at finite temperatures is also examined. In Fig. 7, the mixing enthalpies of the disordered Fe-Co structures are presented for several temperatures: namely low (1 K), intermediate (700 K) and high (1000 K, 1500 K) temperatures. As can be seen, the mixing enthalpies decrease slightly when temperature increases from 1 K to 700 K. At higher temperatures, the variation is more visible as the mixing enthalpies drop faster when temperature varies from 700 K to 1000 K. Interestingly, the trend is inversed as the mixing enthalpies increase with temperature from 1000 K to 1500 K, in the paramagnetic regime. It is also worth noting that the mixing enthalpy curve presents a minimum around 40% Co at 1 K and this minimum slightly shifts to a lower concentration of Co when temperature increases. For all the studied temperatures, the mixing enthalpies remain negative.

In order to better understand these variations, in Fig. 8(a), the mixing enthalpy of several disordered (A2) and an ordered (B2) Fe-Co structures are shown as a function of the temperature. The mixing enthalpies are found to decrease just slightly at low temperatures, below 700 K, and then drop more rapidly in the range from 700 K to 1100 K. Above 1100 K, the enthalpies increase back. As can be observed, all the curves reach a minimum at about 1100 K. In fact, this can be understood as at this point, the energy curves of Fe and Co systems change the curvature as can be seen from the lines with red and blue symbols in the inset of Fig. 8 (b). The change of total energy of each system (pure Fe, pure Co and Fe-50%Co A2) with respect to the respective 1 K value is shown in the main panel of Fig. 8(b). At low temperatures, the change of energy in all the three systems is almost the same, it leads to the almost unchanged mixing enthalpy as defined by Eq. (5). In the range from 700 K to 1100 K, the change in the energies of the pure Fe and pure Co is faster than that in the Fe-Co system due to the strong magnetic disordering near the magnetic phase transition in the pure systems (1065 K for Fe and 1077 K for Co). As a result, the mixing enthalpy decreases (Fig. 8(a)). Above 1100 K, the energies of the pure Fe and Co systems tend to saturate, while the one of the Fe-Co grows up more rapidly as this system goes closer to its Curie point and thus the mixing enthalpy increases back. Similar results are observed in disordered Fe-Co alloys with other Co contents.

## 2.2 B2-A2 phase transition

In this section, the spin-atom MC simulations considering both spin variations and exchanges of atom positions are applied. Such treatment allows us to consider the interplay between the magnetic and the chemical order/disorder, and in particular the role of magnetism ion the chemical phase transition.

### *2.2.1 Mixing tendency of Fe-Co systems*

As revealed by experiments [67] and from the negative mixing enthalpies predicted by DFT and the present model at 0 K (Fig. 2), Fe and Co atoms have a strong mixing and ordering tendency. In order to provide a correct thermodynamic description, we first verify that the coupled spin-atom MC simulations at low temperatures are able to capture these two features of Fe-Co systems.

To observe the mixing tendency via the MC simulations, one of the most convenient starting structures is the Fe-Co bi-phase system as illustrated in Fig. 9 (a). Fig. 9 (b) is a snapshot of the atomic configuration after 400 atom exchange steps and Fig. 9 (c) is the one obtained after 5400 atom exchange steps where Fe and Co atoms are clearly observed to be fully mixed and the disordered A2 phase is formed. After 205400 atom exchange steps (Fig. 9(d)), a highly ordered structure is observed with chem-SRO equal to -0.8024 and 0.8013 for the first and second neighboring shells, respectively which are close to those of the B2 structure (-1 and 1). Thus, the coupled spin-atom equilibration via MC simulations are confirmed to capture correctly the mixing and ordering tendency in Fe-Co structures.

In the following subsections, the spin-atom MC simulations are applied to search for the equilibrium chemical and spin configurations for a given Co concentration at finite temperatures. The chemical order-disorder transition can be determined by following the variation of chemical-LRO as a function of temperature.

### 2.2.2 B2-A2 phase transition within a fixed magnetic state

First, we examine the role of different magnetic states in the B2-A2 ordering phase transition. The spin-atom MC simulations were performed with imposing a magnetic state: either the ferromagnetic, the paramagnetic or the nonmagnetic state. For all temperatures, such states were obtained by constraining the direction of the magnetic moments to be all parallel (FM) or fully random (PM). The nonmagnetic state was obtained by simply setting all the moments equal to zero. We focus on the case of Fe-50%Co.

*Within the ferromagnetic (FM) state*

In Fig. 10, the curve (1) was obtained by constraining the magnetic state at the FM one, which is the magnetic ground state. Within this state, the chemical phase transition temperature $T_{B2-A2}$ is found to be 800 K. It is worth mentioning that the chemical-LRO curves are not well fitted to the power law (Eq. 11). Therefore, the $T_{B2-A2}$ was determined using the inflection point of each curve. This order-disorder transition temperature predicted by the model and MC simulations is found to be lower than the experimental one (about 1000 K) [3]. However, it is worth mentioning that this result is consistent with an estimation based on our DFT data, in which the B2-A2 transition point can be approximately determined by the crossing temperature of the Gibb free energies of formation of B2 and A2 structures with 50% of Co. The DFT results are shown as an inset in Fig. 10 and the crossover is found at 720 K. The DFT calculations were carried out within the perfect ferromagnetic state and the Gibbs free energy of formation is defined as

$$\Delta G^f = \Delta H^f - T.\Delta S^f , \qquad (12)$$

where $\Delta H^f$, $\Delta S^f$ are respectively the enthalpy and the entropy of formation. The latter includes the ideal configurational and the vibrational contributions. Although the crossover temperature is independent of the reference states used, the FM bcc Fe and hcp Co systems were adopted. The calculated vibrational entropies per atom at room temperature for bcc Fe, hcp Co, bcc Fe-Co B2 and A2 (50% Co) are 3.05 $k_B$, 3.33 $k_B$, 3.25 $k_B$, and 3.31 $k_B$, respectively which are in good agreement with experimental values of 3.18 $k_B$ (bcc Fe), 3.27 $k_B$ (hcp Co), 3.18 $k_B$ (B2) and 3.20 $k_B$ (A2 50% Co) [68]. Here $k_B$ is the Boltzmann constant.

Two main features may be at the origin of the rather small difference between the DFT and the model predicted transition temperature (80 K): (i) some chemical disorder (resp. order) can appear in the B2 (resp. A2) phase just below (resp. above) the transition temperature, and (ii) the presence of non-ideal configurational entropy. Both of them are not included in the DFT but in the model + MC determination. On the other side, we propose that the more significant underestimation of the B2-A2 transition temperature by both the DFT and the MC simulations with an imposed ferromagnetic state is mainly because some magnetic entropy, already emerged around 800-1000 K (Fig. 7), are not considered in these approaches.

*Within the paramagnetic (PM) state*

In Fig. 10, the curve (2) was obtained by constraining the magnetic state at a fully paramagnetic one. As shown, the chemical-LRO is found to be visibly lower than 1 even at very low temperatures (near 0 K). This indicates that the Fe-50%Co B2 phase is not stable within the paramagnetic state. By comparing the curves (2) and (1) near 0 K, it suggests that the ferromagnetic order stabilizes the B2 phase at low temperatures and it, in fact, stems from a lower magnetic energy of this phase compared to that of the A2 one at temperatures near 0 K as seen in Fig. 8(a).

*Within the non-magnetic (NM) state*

Another prediction from the model and MC simulations is given for the case of non-magnetic structures. In this case, only the non-magnetic part of the EIM plays a role in the chemical phase transition, that is, with magnetic energy and entropy are missing. The resulting transition temperature is around 1150 K for Fe-50%Co (curve (3)), which is higher than the one obtained by experiments (1000 K). This result is in qualitatively agreement with the previous study by Prasad *et. al.* [42].

### *2.2.3 B2-A2 phase transition with spin equilibration*

All the three curves (1), (2) and (3) in Fig. 10 reveal that the magnetic state and the magneto-chemical coupling strongly influence the chemical transition. Thus, it is necessary to do a full equilibration of both chemical and spin degrees of freedom in order to accurately predict the transition temperature.

The (4)-green curve in Fig. 10 represents the obtained chemical-LRO. The transition temperature $T_{B2-A2}$ is found at 1050 K for the Fe-50%Co alloy, in very good agreement with the experimental data of about 1004 K. To ensure that the obtained transition temperature does not depend on the starting atomic and spin configurations, we performed simulations from the B2 phase at a low temperature (1 K) towards high temperatures, and then from a perfectly random A2 phase at a very high temperature (1900 K) towards low temperatures (blue curve). The slight mismatch of the two curves in Fig. 10 is due to a numerical error.

It is worth mentioning that the present model and other interaction models do commonly not include either explicitly or implicitly the contribution of the lattice vibration. Nevertheless, our model still provides a satisfactory prediction of the transition temperature. This is because in Fe-Co systems, the vibrational entropies of the B2 and the A2 structures are found to be only slightly different. For instance, 3.25 $k_B$, per atom for the B2 and 3.31 $k_B$ for the A2 phase, at 50% Co. Thus, the effect of the lattice vibration can be negligible in the B2-A2 phase transition of the bcc Fe-Co structures.

By carrying out spin-atom equilibrations for other Co concentrations, the transition temperature is also determined as a function of the Co concentration and compared to experiment data [3]. An excellent agreement can be seen in Fig. 11. Together with the predicted Curie temperatures, the present model, although rather simple, is able to predict simultaneously very satisfactory magnetic and chemical phase diagrams for a broad range of temperature and the alloy concentration.

*Magnetism with coupled spin-atom equilibration*

Fig. 12 (a) presents the magnetization with a fully coupled spin-atom equilibration as a function of temperature (blue line). The magnetization curves of the fixed 50%-Co B2 and A2 structures are also shown for comparison. As can be seen from the blue curve in Fig. 10 (a), at the chemical transition temperature $T_{B2-A2}$ = 1050 K, the magnetization is still significantly large. Indeed, at this temperature, Both B2 and A2 structures are still in the ferromagnetic state. As expected, at low temperatures, the blue curve almost coincides with the B2 one while above the order-disorder transition temperature, this curve is well identical to the one of the A2 structure. Interestingly, the second crossover of the magnetization curves of the B2 and A2 phases (see Fig. 5) is identical to the chemical transition temperature obtained from the full spin-atom equilibration.

Mag-SRO is also examined to understand local correlations between magnetic moments. In Fig. 12 (b), the mag-SRO parameters of the Fe-Fe, Fe-Co and Co-Co pairs for the 1st and 2nd neighboring shells are shown. They are all positive (ferromagnetic interaction) for the temperatures considered, even beyond the Curie temperature. The 1st shell Fe-Fe interactions for all the temperatures and the 2nd shell Co-Co interactions below the B2-A2 transition present the highest values. Interestingly, at the chemical transition temperature, all the mag-SRO curves present either a kink or a change of curvature. In addition, the behavior of the 4th shell is similar to the one of the 1st shell, while the behaviour of the 3rd and 5th shells are similar to that of the 2nd shell, for this and all other Co concentrations studied.

## 3. The effect of an external magnetic field on phase transitions

An external magnetic field is widely used as a versatile excitation to tune magnetic properties of materials [69,70]. In this section, the magnetic field is introduced into the EIM to investigate the impact of the field on the magnetic and chemical phase transitions of Fe-Co systems.

In the presence of a magnetic field, the Hamiltonian (Eq. 1) is modified by adding the Zeeman term as

$$H' = H + H_{Zeeman}, \tag{13}$$

with

$$H_{Zeeman} = -g\mu_B \vec{B} \sum_i \vec{M_i}, \tag{14}$$

in which $g$ is the Landé factor and takes the value of 2.0 [31]. Here we consider the case of a constant field.

**3.1 The effect around the zero-field magnetic transition at a given atomic configuration**

First, we consider the effect of the magnetic field around the Curie temperatures of pure bcc Fe and disordered Fe-Co with 50% Co. It is noted that Chui *et al.* [57] have examined this effect for pure bcc Fe using spin-lattice dynamics. In Fig. 13 (a), the magnetization of the pure Fe system under different constant magnetic fields is shown. It is clearly observed that the magnetic field increases the magnetization at all finite temperatures and the effect is stronger at high temperatures, particularly just above the zero-field Curie point. As expected, a stronger field induces a higher magnetization at the same temperature. Such behaviors are consistent with the fact that magnetic fields tend to align the magnetic moments to the direction of the field, thus, they impact more strongly on magnetically disordered than on ferromagnetic states. This observation is similar to what obtained in the spin-lattice dynamics study [57]. We also find a similar effect on the A2 (50% Co) Fe-Co structure as displayed in Fig. 13 (b). The enhanced-ferromagnetism is also seen from the magnetic SRO parameters. Fig. 13 (c) displays the first-neighbor mag-SRO for Fe-Fe, Fe-Co and Co-Co interactions in the Fe-Co A2 structure. It reveals that the mag-SRO is enhanced at all temperatures and the most remarkable effect is observed around the Curie point (1316 K with zero field). The same behavior is also seen for farther neighbor shells and in pure bcc Fe.

The field-induced enhancement of magnetization is found to be almost linear with respect to the increase of the field strength as shown in Fig. 13(d). Interestingly, it is found at all temperatures that with the same field strength, the average moment of the Fe atoms in the Fe-Co system increases more strongly than that of the Co atoms in the same systems, as can be observed from the lines with square (Fe) and triangle (Co) symbols in Fig. 13 (d). This result may be associated to the electronic properties of Fe and Co atoms in pure and Fe-Co systems where Co has an essentially saturated magnetism, while Fe can still enhance its magnetism in the presence of excitations, by for example, shifting down the highest peak in the majority-spin states of Fe atoms towards the Fermi level in Fe-Co structures [43].

It is also worth noting that, below the Curie temperature of pure Fe (e.g. at 1000 K), Fe moments in pure Fe increase faster than those in the Fe-Co systems with increasing field strength, as seen in Fig. 13 (d), from the dot black line with the filled diamond symbols. This feature can be understood as the field induces a stronger impact on the more disorderly magnetic system and it is also consistent with the magnetic disordering in pure Fe and Fe-Co observed in Figs. 13 (a), (b) and (c).

In contrast, at high temperatures near the Curie point of the Fe-Co systems, e.g. at 1500 K the enhancement in the moment of the Fe atoms in pure bcc Fe is slower than that of the counterpart in the Fe-Co systems, which is reflected by the solid black line with the filled diamond symbols in Fig. 13 (d).

### 3.2 The effect on the B2-A2 chemical transition temperature

As shown above in Fig. 10, magnetism plays a vital role in the order-disorder transition temperature $T_{B2-A2}$. While a magnetic field has a significant influence on the magnetic order of the structure. Therefore, it is worth investigating the effect of magnetic fields on the chemical phase transition in Fe-Co alloys.

To this end, we performed MC simulations for Fe-50%Co systems, starting with the B2 structure at low temperatures. Both the spin and chemical configurations are equilibrated at each temperature in the presence of a magnetic field. The resulting chemical long-range-order evolution is presented in Fig. 14. Interestingly, the chemical transition is almost unaltered by the considered fields. This observation is similar to the finding in Fe-Ni alloys subjected to a magnetic field [31]. In fact, this result can be understood as the order-disorder transition temperature (1050 K) is sufficiently lower than the Curie point, where the effect of the magnetic fields on the magnetization is weak (Fig. 13 (b)).

## V. Conclusion

We have proposed an *ab-initio* based effective interaction model for the study of magnetism, thermodynamics, and their interplay in body-centered cubic Fe-Co alloys, with Co content from 0 to 70%. This model consists in a pairwise magnetic and a chemical-interaction term. For the former, a Heisenberg formalism is adopted. But; we go beyond the conventional formalism to account for the variation of the spin magnitude of each Fe atom as a function of its local chemical environment, as predicted by DFT.

The current model has successfully reproduced the energetic landscape and the ground-state magnetic moments of both ordered (B2) and disordered (A2) Fe-Co systems, as obtained by DFT. Moreover, in combination with atomic Monte Carlo simulations, the obtained Curie and chemical B2-A2 phase transition temperatures were found to be in good agreement with available experimental data for a wide range of concentration (0 to 70% Co). A strong dependency of the chemical transition temperature on the magnetic configuration is evidenced. In particular, the stability of the B2 phase at low and intermediate temperatures are clearly favored by the ferromagnetism. In addition, we pointed out a much more important effect of magnetic rather than vibrational entropy on the chemical transition. The above-mentioned findings attest the ability of this model to predict other magneto-energetic properties in stable and metastable bcc Fe-Co structures at any given temperature, which is useful to compare with and to interpret experimental results.

Furthermore, as expected, we found a significant effect of an external magnetic field to enhance the magnetization and the magnetic short-range orders, even below the Curie point, However, the B2-A2 transition temperatures are unaltered with fields with a commonly accessible strength. This finding is similar to the case of order-disorder transition in fcc Fe-Ni alloys.

## Acknowledgments

This work was supported by the French-German ANR-DFG MAGIKID project. The DFT calculations were performed using the DARI-GENCI CPU resources within the A0050906020 project, and the CINECA-MARCONI supercomputer within the SISTEEL project.

## References

[1]   Sourmail T 2005 Near equiatomic FeCo alloys: Constitution, mechanical and magnetic properties *Prog. Mater. Sci.* **50** 816–80

[2]   Abbasov R M, Mekhrabov A O, Mustafaev R A and Binnatov K G 1976 A calorimetric study of lead-thallium alloys *Sov. Phys. J.* **19** 4–6

[3]   Massalski T B and Kamoto H 1990 *Binary alloy phase diagrams* (ASM International)

[4]   Elmen G W 1929 Magnetic Alloys of Iron, Nickel, and Cobalt *Bell Syst. Tech. J.* **8** 435–65

[5]   MacLaren J M, Schulthess T C, Butler W H, Sutton R and McHenry M 1999 Electronic structure, exchange interactions, and Curie temperature of FeCo *J. Appl. Phys.* **85** 4833–5

[6]   Neumayer M and Fähnle M 2001 Atomic defects in FeCo: Stabilization of the (formula presented) structure by magnetism *Phys. Rev. B - Condens. Matter Mater. Phys.* **64** 2–5

[7]   Mizuno M, Araki H and Shirai Y 2006 First Principles Calculation of Defect and Magnetic Structures in FeCo *Mater. Trans.* **47** 2646–50

[8]   Alf D, Gillan M J and Price G D 2007 Temperature and composition of the Earth's core

*Contemp. Phys.* **48** 63–80

[9]   Wu D, Zhang Q, Liu J P, Yuan D and Wu R 2008 First-principles prediction of enhanced magnetic anisotropy in FeCo alloys *Appl. Phys. Lett.* **92** 2008–10

[10]  Rahaman M, Ruban A V., Mookerjee A and Johansson B 2011 Magnetic state effect upon the order-disorder phase transition in Fe-Co alloys: A first-principles study *Phys. Rev. B - Condens. Matter Mater. Phys.* **83** 1–8

[11]  Cuong Nguyen M, Zhao X, Ji M, Wang C Z, Harmon B and Ho K M 2012 Atomic structure and magnetic properties of Fe 1-xCo x alloys *J. Appl. Phys.* **111** 07E338

[12]  Messina L, Nastar M, Garnier T, Domain C and Olsson P 2014 Exact ab initio transport coefficients in bcc Fe − X ( X = Cr , Cu , Mn , Ni , P , Si ) dilute alloys *Phys. Rev. B* **90** 104203

[13]  Ding J, Zhang P, Li X, Wang Y, Huang S and Zhao J 2017 Magnetism and energetics for vacancy and helium impurity in Fe-9Cr alloy: A first-principles study *Comput. Mater. Sci.* **138** 267–76

[14]  Lizárraga R, Pan F, Bergqvist L, Holmström E, Gercsi Z and Vitos L 2017 First Principles Theory of the hcp-fcc Phase Transition in Cobalt *Sci. Rep.* **7** 6–13

[15]  Messina L, Nastar M, Sandberg N and Olsson P 2016 Systematic electronic-structure investigation of substitutional impurity diffusion and flux coupling in bcc iron *Phys. Rev. B* **93** 1–18

[16]  Versteylen C D, Van Dijk N H and Sluiter M H F 2017 First-principles analysis of solute diffusion in dilute bcc Fe- X alloys *Phys. Rev. B* **96** 1–9

[17]  Soisson F and Fu C C 2007 Cu-precipitation kinetics in α-Fe from atomistic simulations: Vacancy-trapping effects and Cu-cluster mobility *Phys. Rev. B - Condens. Matter Mater. Phys.* **76** 1–12

[18]  Arroyo-De Dompablo M E, Morales-Garca A and Taravillo M 2011 DFTU calculations of crystal lattice, electronic structure, and phase stability under pressure of TiO2 polymorphs *J. Chem. Phys.* **135**

[19]  Jain A, Hautier G, Ong S P, Moore C J, Fischer C C, Persson K A and Ceder G 2011 Formation enthalpies by mixing GGA and GGA + U calculations *Phys. Rev. B - Condens. Matter Mater. Phys.* **84** 045115

[20]  Dang M and Rancourt D 1996 Simultaneous magnetic and chemical order-disorder

phenomena in Fe3Ni, FeNi, and FeNi3 *Phys. Rev. B - Condens. Matter Mater. Phys.* **53** 2291–302

[21] Ackland G J 2009 Ordered sigma-type phase in the Ising model of Fe-Cr stainless steel *Phys. Rev. B - Condens. Matter Mater. Phys.* **79** 1–6

[22] Paduani C, Plascak J A and Galv Da Silva E 1991 Simple Ising model for the magnetic properties of Fe-Mn alloys *Phys. Rev. B* **44** 9715–8

[23] Freitas A S, De Albuquerque D F and Moreno N O 2014 Fe-Mn alloys: A mixed-bond spin-1/2 Ising model version *J. Magn. Magn. Mater.* **361** 137–9

[24] Shi X, Zhao J and Xu X 2015 Phase diagram of the mixed Ising model with Fe4N structure under a time-dependent oscillating magnetic field *Phys. A Stat. Mech. its Appl.* **419** 234–40

[25] Salazar M, Zamora L E, Pérez Alcazar G A, Plascak J A and Aguirre W R 2002 Site-diluted quantum Heisenberg spin model applied to the magnetic properties of Fe-Al disordered alloys *Phys. B Condens. Matter* **320** 236–8

[26] Aguirre Contreras W R, Zamora L E, Pérez Alcázar G A and Plascak J A 2005 Magnetic properties of Fe-Al alloys through Ising and Heisenberg models with a nonlinear exchange interaction *Phys. Rev. B - Condens. Matter Mater. Phys.* **72** 1–4

[27] Ležaić M, Mavropoulos P and Blügel S 2007 First-principles prediction of high Curie temperature for ferromagnetic bcc-Co and bcc-FeCo alloys and its relevance to tunneling magnetoresistance *Appl. Phys. Lett.* **90** 0–3

[28] Blizak S, Bihlmayer G and Blügel S 2012 Ab initio investigations of magnetic properties of FeCo monolayer alloy films on Rh(001) *Phys. Rev. B - Condens. Matter Mater. Phys.* **86** 1–8

[29] Tahir-Kheli R A and Kawasaki T 1977 Simultaneous occurrence of magnetic and spatial long-range order in binary alloys *J. Phys. C Solid State Phys.* **10** 2207–21

[30] Sanchez J M and Lin C H 1984 Modeling of magnetic and chemical ordering in binary alloys *Phys. Rev. B* **30** 1448–53

[31] Vernyhora I V., Ledue D, Patte R and Zapolsky H 2010 Monte Carlo investigation of the correlation between magnetic and chemical ordering in NiFe alloys *J. Magn. Magn. Mater.* **322** 2465–70

[32] Ekholm M, Zapolsky H, Ruban A, Vernyhora I, Ledue D and Abrikosov I A 2010 Influence of the magnetic state on the chemical order-disorder transition temperature in Fe-Ni22

permalloy. *Phys. Rev. Lett.* **105** 167208

[33] Alés A and Lanzini F 2014 Atomic and magnetic ordering in bcc Cu-Al-Mn: Computational study *Model. Simul. Mater. Sci. Eng.* **22**

[34] Lavrentiev M Y, Dudarev S L and Nguyen-Manh D 2009 Magnetic cluster expansion simulations of FeCr alloys *J. Nucl. Mater.* **386–388** 22–5

[35] Lavrentiev M Y, Soulairol R, Fu C C, Nguyen-Manh D and Dudarev S L 2011 Noncollinear magnetism at interfaces in iron-chromium alloys: The ground states and finite-temperature configurations *Phys. Rev. B - Condens. Matter Mater. Phys.* **84** 1–13

[36] Lavrentiev M Y, Wróbel J S, Nguyen-Manh D, Dudarev S L, Ganchenkova M G, Yu Lavrentiev M, Wróbel J S, Nguyen-Manh D, Dudarev S L, Ganchenkova M G and Wr obel J S 2016 Magnetic cluster expansion model for random and ordered magnetic face-centered cubic Fe-Ni-Cr alloys *J. Appl. Phys.* **120** 7–123

[37] Lavrentiev M Y, Dudarev S L and Nguyen-Manh D 2011 Magnetic cluster expansion model for high-temperature magnetic properties of iron and iron-chromium alloys *J. Appl. Phys.* **109** 07E123

[38] Lavrentiev M Y, Nguyen-Manh D and Dudarev S L 2010 Magnetic cluster expansion model for bcc-fcc transitions in Fe and Fe-Cr alloys *Phys. Rev. B* **81** 184202

[39] Lavrentiev M Y, Wróbel J S, Nguyen-Manh D and Dudarev S L 2014 Magnetic and thermodynamic properties of face-centered cubic Fe-Ni alloys *Phys. Chem. Chem. Phys.* **16** 16049–59

[40] Lavrentiev M Y, Nguyen-Manh D and Dudarev S L 2010 Cluster expansion models for Fe-Cr alloys, the prototype materials for a fusion power plant *Comput. Mater. Sci.* **49** S199–203

[41] Pierron-Bohnes V, Cadeville M C and Gautier F 1983 Magnetism and local order in dilute FeCo alloys *J. Phys. F Met. Phys.* **13** 1689–713

[42] Prasad A, Sanyal B and Mookerjee A 2014 Journal of Magnetism and Magnetic Materials Study of the effect of magnetic ordering on order – disorder transitions in binary alloys *J. Magn. Magn. Mater.* **360** 15–20

[43] Tran V T, Fu C C and Li K 2020 Predicting magnetization of ferromagnetic binary Fe alloys from chemical short range order *Comput. Mater. Sci.* **172** 109344

[44] Collins M F and Forsyth J B 1963 The magnetic moment distribution in some transition

metal alloys *Philos. Mag.* **8** 401

[45] Hasegawa H 1983 A spin fluctuation theory of degenerate narrow bands-finite-temperature magnetism of iron *J. Phys. F Met. Phys.* **13** 1915–29

[46] Kakehashi Y and Atiqur R. Patoary M 2011 First-Principles Dynamical Coherent-Potential Approximation Approach to the Ferromagnetism of Fe, Co, and Ni *J. Phys. Soc. Japan* **80** 034706

[47] Amara H, Fu C C, Soisson F and Maugis P 2010 Aluminum and vacancies in α-iron: Dissolution, diffusion, and clustering *Phys. Rev. B - Condens. Matter Mater. Phys.* **81** 1–11

[48] Kresse G and Hafner J 1993 Ab initio molecular dynamics for liquid metals *Phys. Rev. B* **47** 558–61

[49] Kresse G and Furthmüller J 1996 Efficiency of ab-initio total energy calculations for metals and semiconductors using a plane-wave basis set *Comput. Mater. Sci.* **6** 15–50

[50] Kresse G and Furthmüller J 1996 Efficient iterative schemes for ab initio total-energy calculations using a plane-wave basis set *Phys. Rev. B - Condens. Matter Mater. Phys.* **54** 11169–86

[51] Kresse G and Joubert D 1999 From ultrasoft pseudopotentials to the projector augmented-wave method *Phys. Rev. B* **59** 1758–75

[52] Perdew J P, Burke K and Ernzerhof M 1996 Generalized Gradient Approximation Made Simple *Phys. Rev. Lett.* **77** 3865–8

[53] Monkhorst H J and Pack J D 1976 Special points for Brillouin-zone integrations *Phys. Rev. B* **13** 5188–92

[54] Lavrentiev M Y and Dudarev S L 2009 Magnetic cluster expansion simulations of FeCr alloys *J. Nucl. Mater.* **386**–**388** 22–5

[55] Cowley J M 1950 An approximate theory of order in alloys *Phys. Rev.* **77** 669–75

[56] Martinez E, Fu C C, Levesque M, Nastar M and Soisson F 2011 Simulations of Decomposition Kinetics of Fe-Cr Solid Solutions during Thermal Aging *Solid State Phenom.* **172**–**174** 1016–21

[57] Chui C P and Zhou Y 2014 Investigating the magnetovolume effect in isotropic body-centered-cubic iron using spin-lattice dynamics simulations *AIP Adv.* **4** 087123

[58] Xie Y, Fan J, Xu L, Zhang X, Xu R, Zhu Y, Tang R, Wang C, Ma C, Pi L, Zhang Y and Yang H 2019 Unambiguous determining the Curie point in perovskite manganite with

second-order phase transition by scaling method *Phys. Lett. A* **383** 125843

[59]   Connelly D L, Loomis J S and Mapother D E 1971 Specific Heat of Nickel near the Curie Temperature *Phys. Rev. B* **3** 924–34

[60]   Carvalho T T, Fernandes J R A, Perez De La Cruz J, Vidal J V., Sobolev N A, Figueiras F, Das S, Amaral V S, Almeida A, Agostinho Moreira J and Tavares P B 2013 Room temperature structure and multiferroic properties in Bi 0.7La0.3FeO3 ceramics *J. Alloys Compd.* **554** 97–103

[61]   Yang Z, Gao D, Zhang J, Xu Q, Shi S, Tao K and Xue D 2015 Realization of high Curie temperature ferromagnetism in atomically thin MoS2 and WS2 nanosheets with uniform and flower-like morphology *Nanoscale* **7** 650–8

[62]   Fuh H-R, Chang C-R, Wang Y-K, Evans R F L L, Chantrell R W and Jeng H-T 2016 Newtype single-layer magnetic semiconductor in transition-metal dichalcogenides VX2 (X = S, Se and Te) *Sci. Rep.* **6** 32625

[63]   Bozorth R M 1951 *Ferromagnetism* (New York : Van Nostrand)

[64]   Jacobs I S, Patchen H J and Johnson N A 1991 Raising the Curie point in the iron-cobalt-(aluminum) system *J. Appl. Phys.* **69** 5924–6

[65]   Hansen M 1958 *Constitution of Binary Alloys* (New York: McGraw-Hill)

[66]   Forrer R 1930 Le problème des deux points de Curie *J. Phys. le Radium* **1** 49–64

[67]   Wang H, Lück R and Predel B 1993 Mixing enthalpy of liquid Fe-Co-Ti alloys *J. Non. Cryst. Solids* **156**–**158** 388–92

[68]   Lucas M S, Munz J A, Mauger L, Li C W, Sheets A O, Turgut Z, Horwath J, Abernathy D L, Stone M B, Delaire O, Xiao Y and Fultz B 2010 Effects of chemical composition and B2 order on phonons in bcc Fe-Co alloys *J. Appl. Phys.* **108** 1–6

[69]   Alzate-Cardona J D, Barco-Ríos H and Restrepo-Parra E 2018 Dynamic phase transitions in La2/3 Ca1/3 MnO3 manganites: A Monte Carlo simulation study *Phys. Lett. A* **382** 792–7

[70]   Shen S Q and Zhang F C 2002 Antiferromagnetic Heisenberg model on an anisotropic triangular lattice in the presence of a magnetic field *Phys. Rev. B - Condens. Matter Mater. Phys.* **66** 1–4

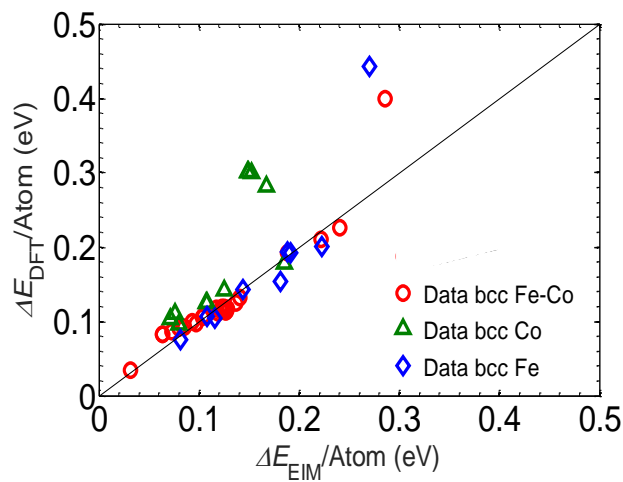

**Fig. 1**: Energy difference between two systems with different magnetic configurations but the same atomic structure: comparison between DFT and EIM results. For each system, the EIM adopted the DFT predicted magnetic moments.

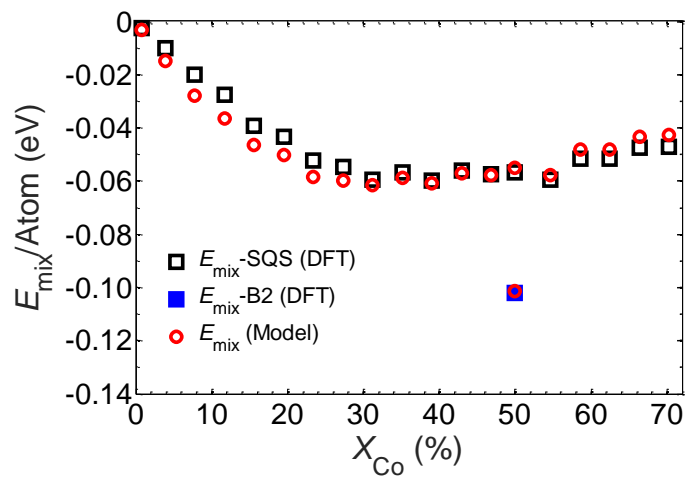

**Fig. 2**: Mixing enthalpies obtained by the EIM compared with the DFT data. For each system, the EIM adopted the DFT predicted magnetic moments.

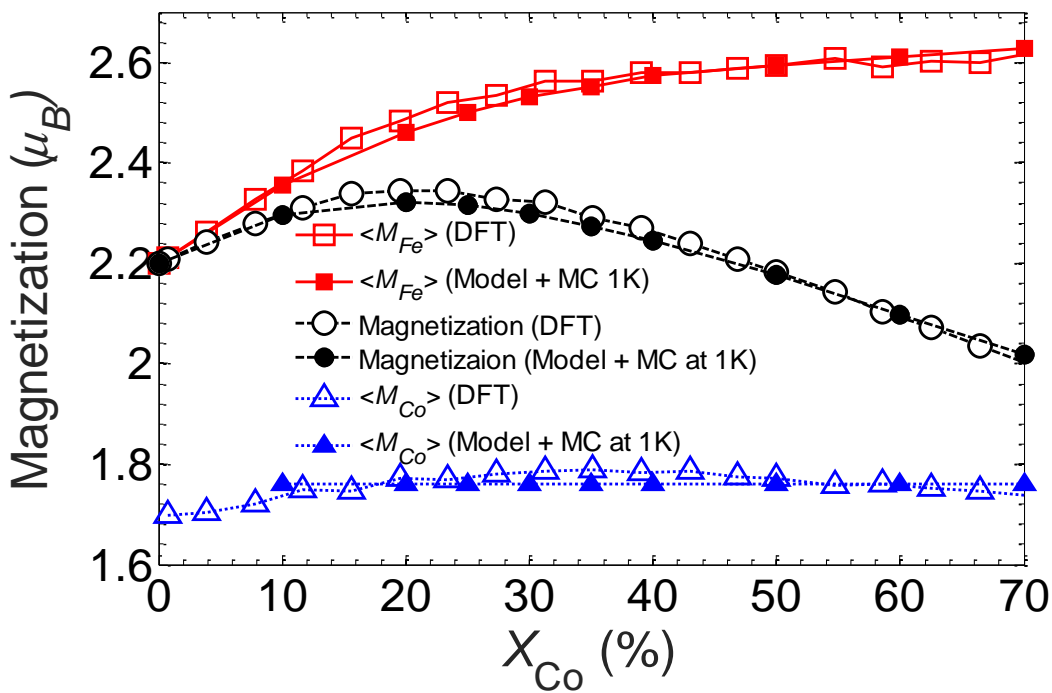

**Fig. 3**: Magnetization of disordered structures obtained by the model (EIM) + MC simulations at 1 K, and compared with the DFT data.

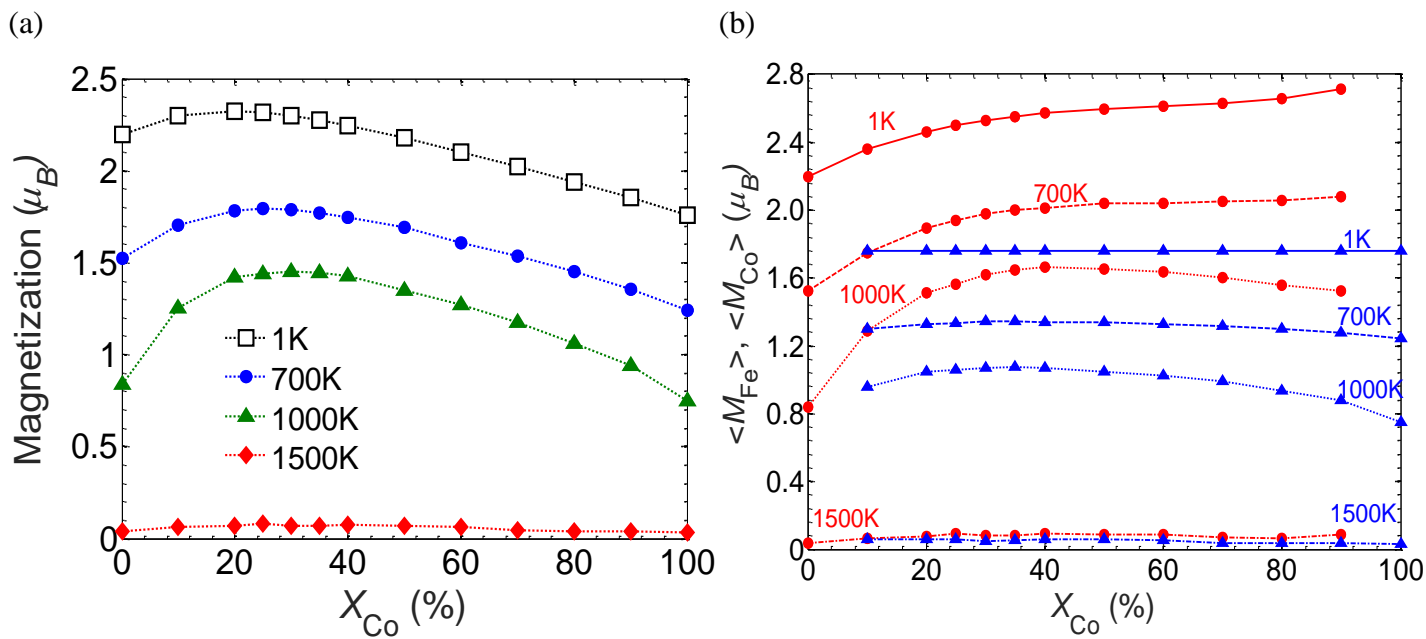

**Fig. 4**: (a) Magnetization and (b) average magnetic moments of each species at various temperatures as predicted by the EIM + MC simulations.

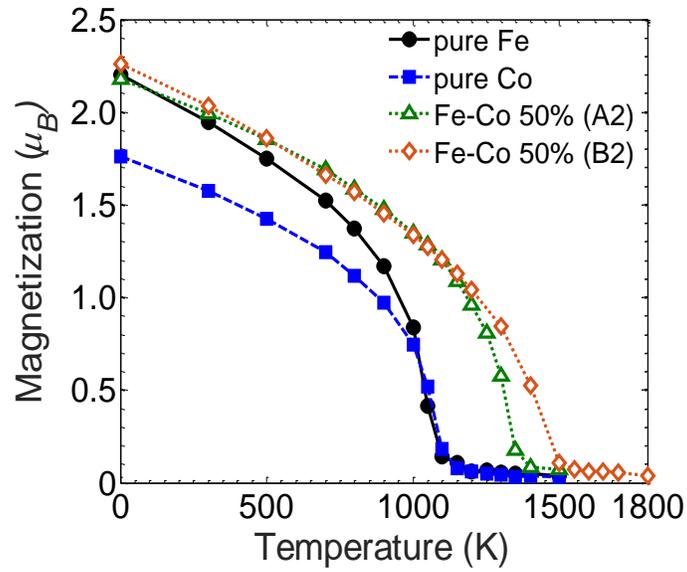

**Fig. 5**: Magnetization as a function of temperature for different systems: pure bcc Fe (black), pure bcc Co (blue), bcc Fe-50%Co A2 (green) and B2 (orange).

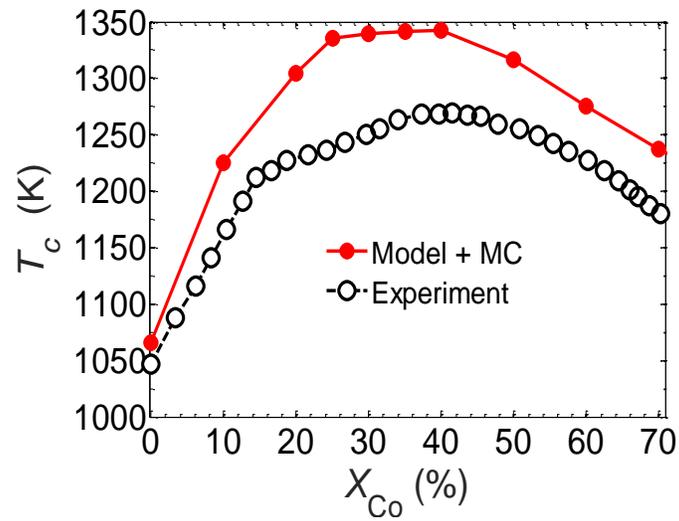

**Fig. 6**: Curie temperature of the A2 phase as a function of Co concentration: the EIM + MC simulations versus experimental data [64].

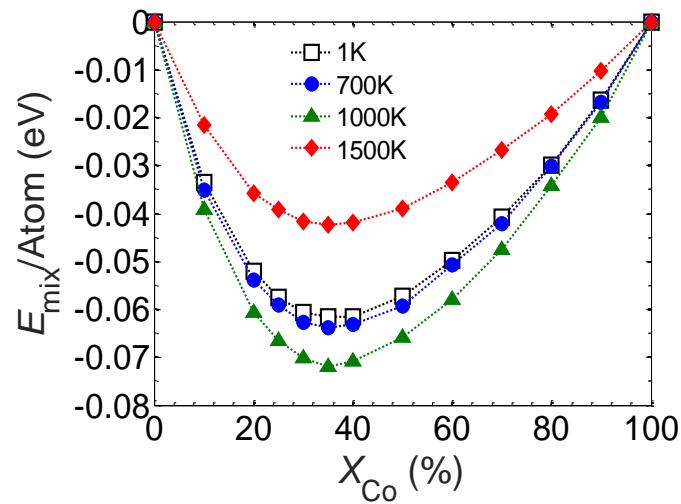

**Fig. 7**: Mixing enthalpy as a function of concentration predicted for several temperatures by the EIM + MC simulations.

(a) (b)

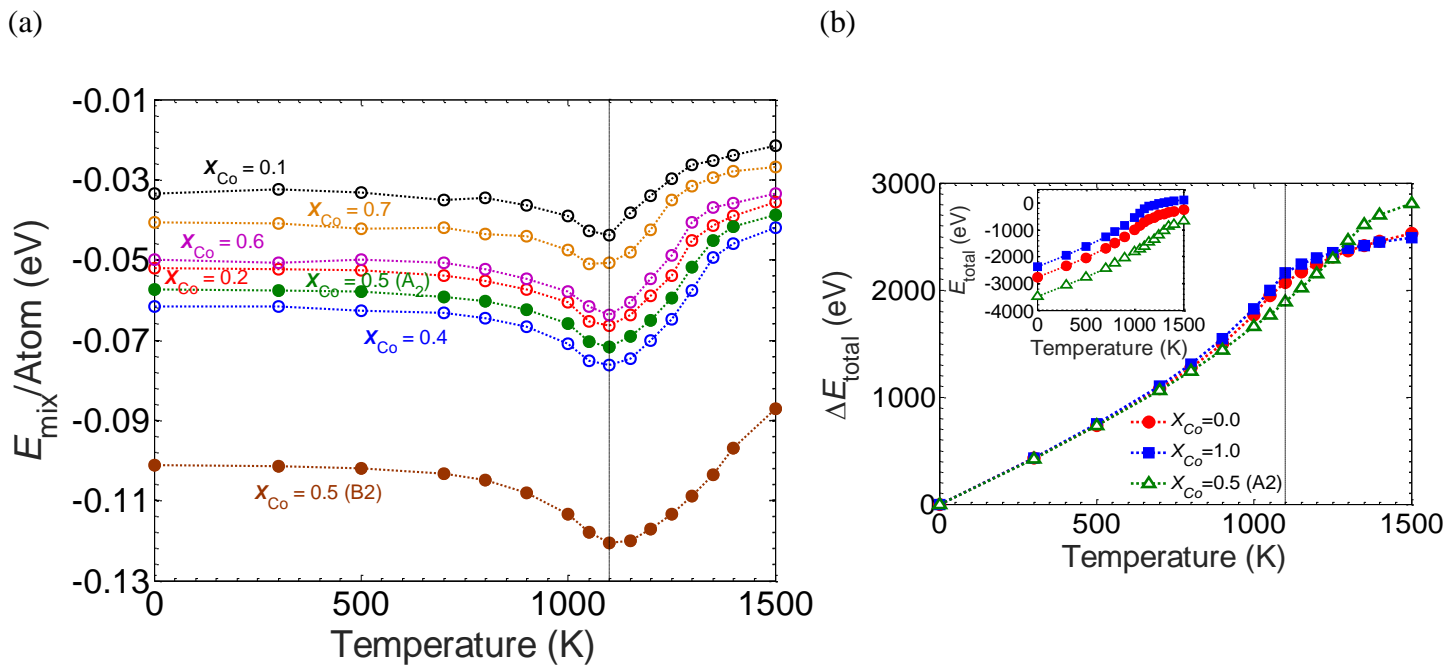

**Fig. 8**: (a) Mixing enthalpy as a function of temperature. (b) Change of total energy relative to the respective 1 K value in the pure Fe, pure Co and Fe-Co (A2 50% Co) phases. The inset presents the total energies of these systems.

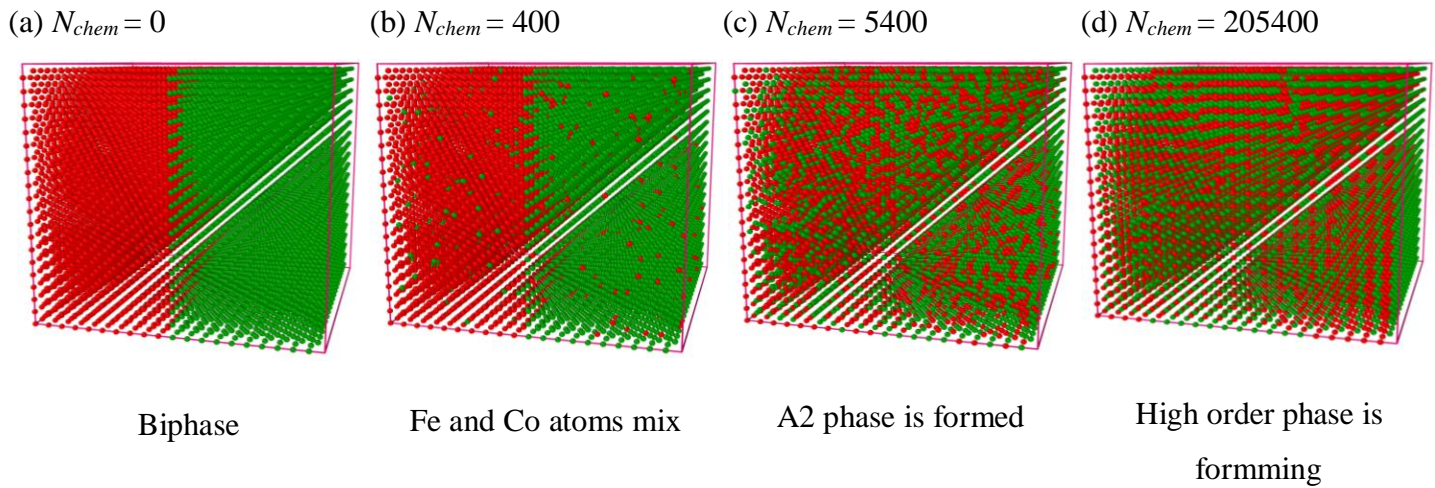

(a) $N_{chem} = 0$    (b) $N_{chem} = 400$    (c) $N_{chem} = 5400$    (d) $N_{chem} = 205400$

Biphase    Fe and Co atoms mix    A2 phase is formed    High order phase is formming

**Fig. 9**: Mixing tendency of Fe-Co alloys observed in the 1K MC simulations using the EIM: snapshots obtained with the full spin-atom equilibration.

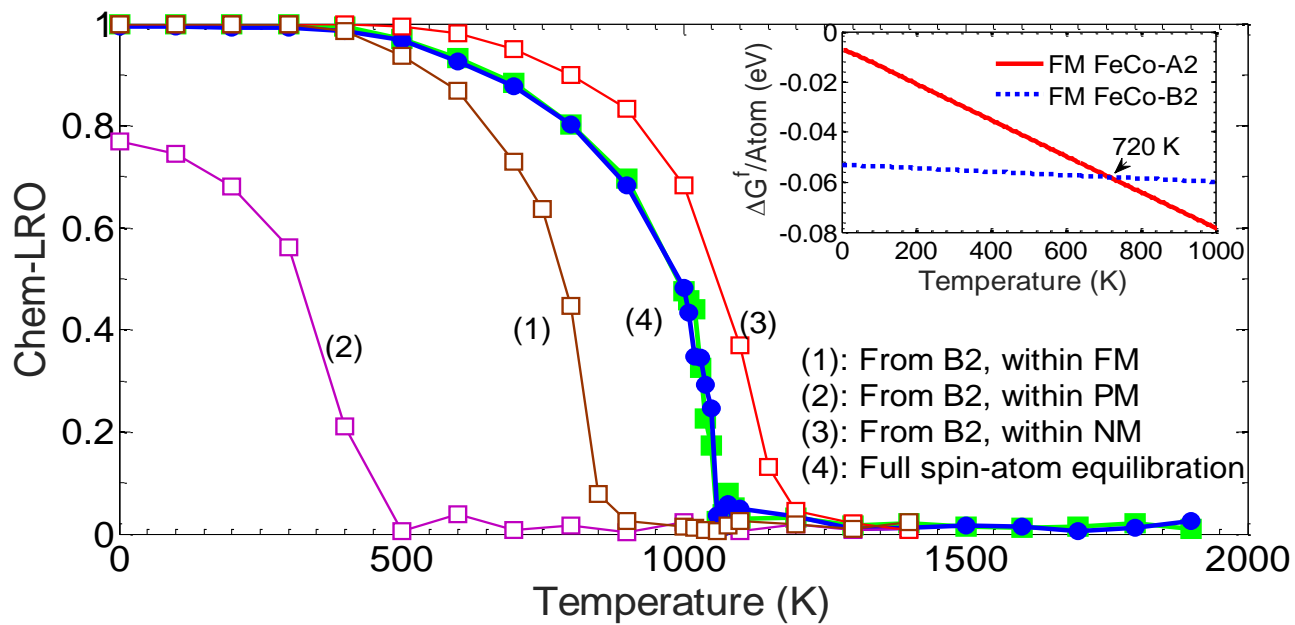

(1): From B2, within FM
(2): From B2, within PM
(3): From B2, within NM
(4): Full spin-atom equilibration

**Fig. 10**: Chem-LRO as a function of temperature obtained by coupled spin-atom equilibrations within different imposed magnetic states (curves 1, 2, 3) and by the full spin-atom equilibration (curves 4).

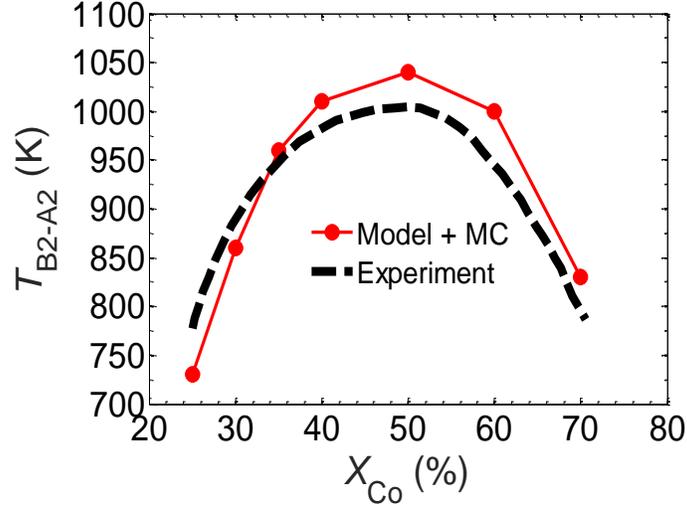

**Fig. 11**: B2-A2 transition temperatures obtained for different Co concentrations: the EIM + MC simulations versus experimental data [3].

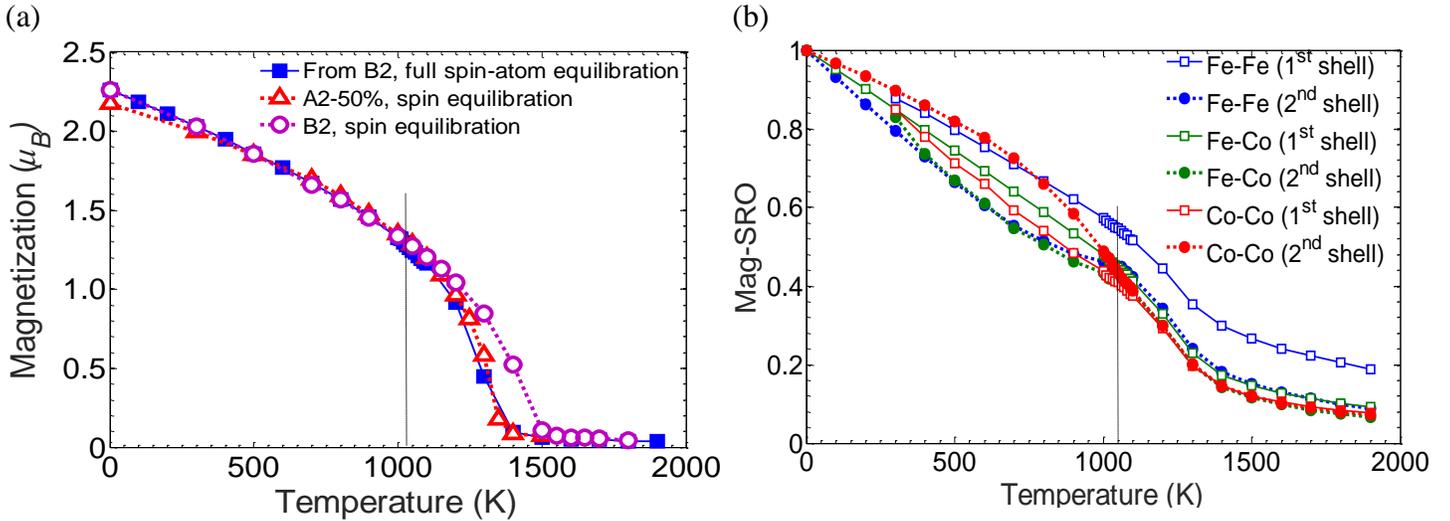

**Fig. 12**: (a) Magnetization obtained from the fully coupled spin-atom equilibration, starting from the B2 phase (blue), compared with the magnetization of the B2 (violet) and the A2 (red) phase with spin equilibration. (b) $1^{st}$ and $2^{nd}$ shell mag-SRO for Fe-Fe, Fe-Co and Co-Co pairs. The vertical dash lines indicate the B2-A2 transition temperature.

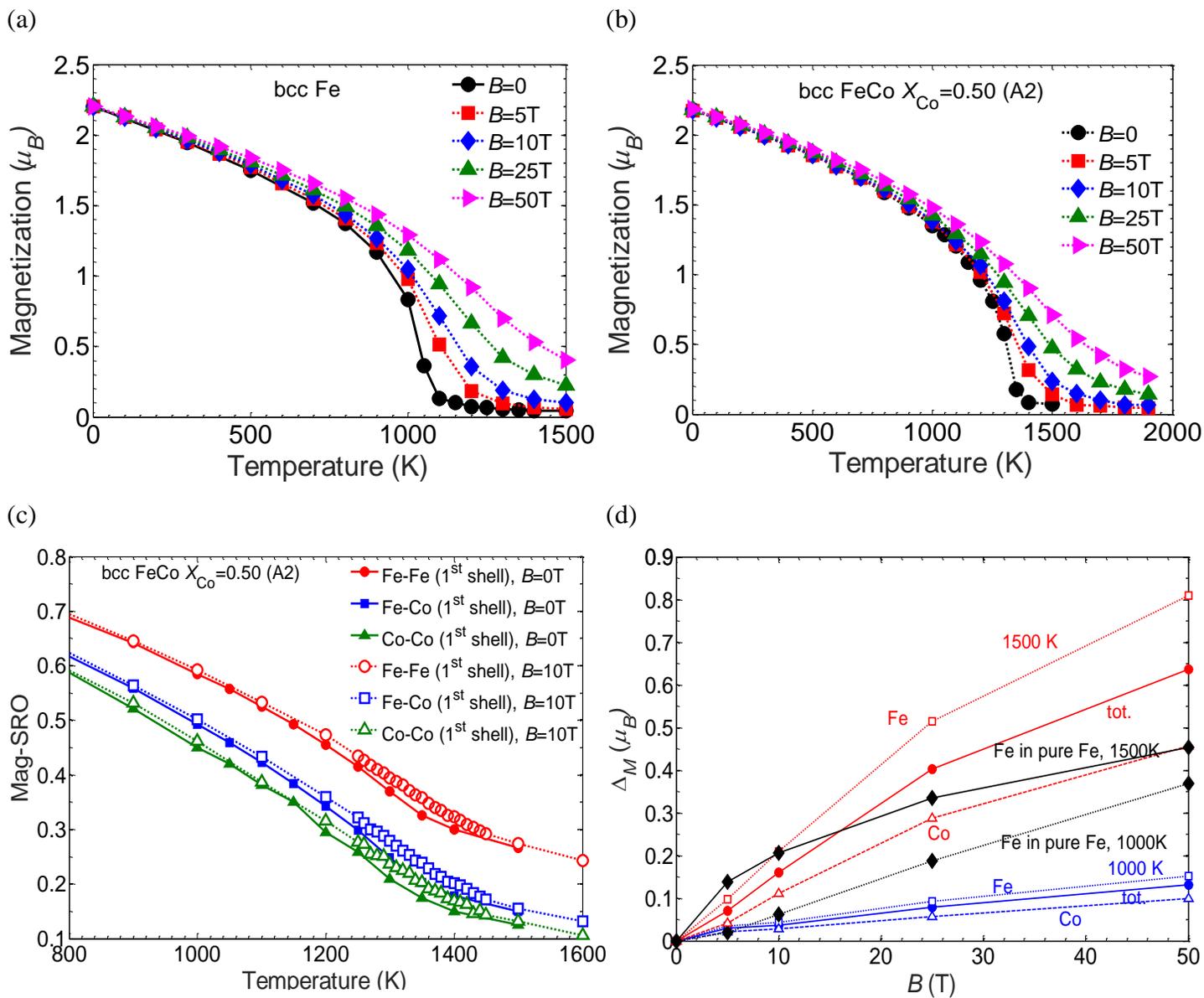

**Fig. 13**: Magnetization as a function of temperature in the presence of the magnetic field in (a) pure bcc Fe, (b) bcc Fe-50%Co A2. (c) Mag-SRO as a function of temperature under different magnetic fields. (d) The change in the average moments of Fe (square) and Co (triangular) atoms and magnetization (circles denoted as "tot") as a function of the field strength.

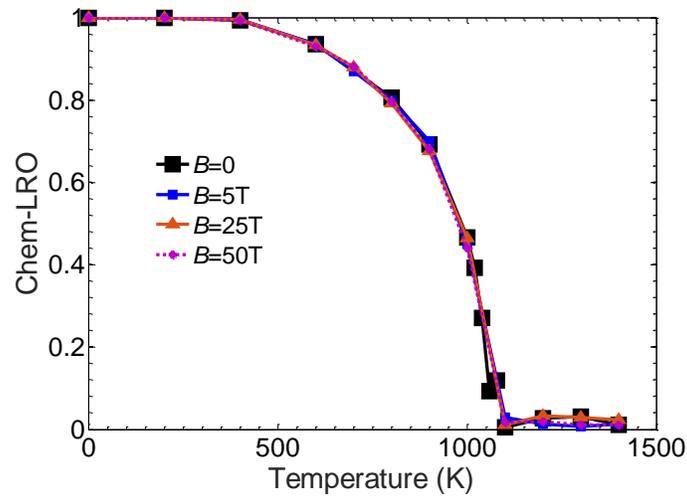

**Fig. 14**: Chemical-LRO of the Fe-50%Co system as a function of temperature under the effect of different external magnetic fields. The results were obtained with the fully coupled spin-atom equilibration.

**Table I**: Fitted parameters (in meV) for the magnetic ($J_{ij}$) and chemical ($V_{ij}$) pair interactions in the EIM (Eq. 1)

|  | 1nn | 2nn | 3nn | 4nn | 5nn |
|---|---|---|---|---|---|
| $J_{Fe-Fe}$ | -11.65800 | -1.67630 | 0.51512 | 0.49840 | 0.22120 |
| $J_{Co-Co}$ | -7.60200 | -6.58240 | -5.16990 | 1.95370 | 0.13135 |
| $J_{Fe-Co}$ | -5.98790 | -2.69460 | -2.00000 | -0.04000 | 0.01000 |
| $V_{Fe-Fe}$ | 10.62000 | -4.94300 | -- | -- | -- |
| $V_{Co-Co}$ | 10.62000 | -4.94300 | -- | -- | -- |
| $V_{Fe-Co}$ | -11.92500 | 4.10800 | -- | -- | -- |